\documentclass[journal,twoside,twocolumn]{IEEEtran}
\IEEEoverridecommandlockouts
\usepackage{cite}
\usepackage[T1]{fontenc}
\usepackage{graphicx}
\usepackage{amssymb}
\usepackage{amsmath}
\usepackage{subfigure}
\usepackage{booktabs} 
\usepackage{amsthm}
\usepackage{microtype} 
\usepackage{tablefootnote}
\usepackage{hyperref}
\usepackage{balance}
\usepackage{color}

\usepackage{amssymb}
\usepackage{amsfonts}
\usepackage{mathrsfs}
\usepackage{xspace}
\usepackage{bm}
\usepackage{upgreek}

\newcommand{\safemath}[2]{\newcommand{#1}{\ensuremath{#2}\xspace}}

\safemath{\bma}{\mathbf{a}}
\safemath{\bmb}{\mathbf{b}}
\safemath{\bmc}{\mathbf{c}}
\safemath{\bmd}{\mathbf{d}}
\safemath{\bme}{\mathbf{e}}
\safemath{\bmf}{\mathbf{f}}
\safemath{\bmg}{\mathbf{g}}
\safemath{\bmh}{\mathbf{h}}
\safemath{\bmi}{\mathbf{i}}
\safemath{\bmj}{\mathbf{j}}
\safemath{\bmk}{\mathbf{k}}
\safemath{\bml}{\mathbf{l}}
\safemath{\bmm}{\mathbf{m}}
\safemath{\bmn}{\mathbf{n}}
\safemath{\bmo}{\mathbf{o}}
\safemath{\bmp}{\mathbf{p}}
\safemath{\bmq}{\mathbf{q}}
\safemath{\bmr}{\mathbf{r}}
\safemath{\bms}{\mathbf{s}}
\safemath{\bmt}{\mathbf{t}}
\safemath{\bmu}{\mathbf{u}}
\safemath{\bmv}{\mathbf{v}}
\safemath{\bmw}{\mathbf{w}}
\safemath{\bmx}{\mathbf{x}}
\safemath{\bmy}{\mathbf{y}}
\safemath{\bmz}{\mathbf{z}}
\safemath{\bmzero}{\mathbf{0}}
\safemath{\bmone}{\mathbf{1}}

\bmdefine{\biad}{a}
\bmdefine{\bibd}{b}
\bmdefine{\bicd}{c}
\bmdefine{\bidd}{d}
\bmdefine{\bied}{e}
\bmdefine{\bifd}{f}
\bmdefine{\bigd}{g}
\bmdefine{\bihd}{h}
\bmdefine{\biid}{i}
\bmdefine{\bijd}{j}
\bmdefine{\bikd}{k}
\bmdefine{\bild}{l}
\bmdefine{\bimd}{m}
\bmdefine{\bind}{n}
\bmdefine{\biod}{o}
\bmdefine{\bipd}{p}
\bmdefine{\biqd}{q}
\bmdefine{\bird}{r}
\bmdefine{\bisd}{s}
\bmdefine{\bitd}{t}
\bmdefine{\biud}{u}
\bmdefine{\bivd}{v}
\bmdefine{\biwd}{w}
\bmdefine{\bixd}{x}
\bmdefine{\biyd}{y}
\bmdefine{\bizd}{z}

\bmdefine{\bixid}{\xi}
\bmdefine{\bilambdad}{\lambda}
\bmdefine{\bimud}{\mu}
\bmdefine{\bithetad}{\theta}
\bmdefine{\biphid}{\phi}
\bmdefine{\bideltad}{\delta}

\safemath{\bmia}{\biad}
\safemath{\bmib}{\bibd}
\safemath{\bmic}{\bicd}
\safemath{\bmid}{\bidd}
\safemath{\bmie}{\bied}
\safemath{\bmif}{\bifd}
\safemath{\bmig}{\bigd}
\safemath{\bmih}{\bihd}
\safemath{\bmii}{\biid}
\safemath{\bmij}{\bijd}
\safemath{\bmik}{\bikd}
\safemath{\bmil}{\bild}
\safemath{\bmim}{\bimd}
\safemath{\bmin}{\bind}
\safemath{\bmio}{\biod}
\safemath{\bmip}{\bipd}
\safemath{\bmiq}{\biqd}
\safemath{\bmir}{\bird}
\safemath{\bmis}{\bisd}
\safemath{\bmit}{\bitd}
\safemath{\bmiu}{\biud}
\safemath{\bmiv}{\bivd}
\safemath{\bmiw}{\biwd}
\safemath{\bmix}{\bixd}
\safemath{\bmiy}{\biyd}
\safemath{\bmiz}{\bizd}

\safemath{\bmxi}{\bixid}
\safemath{\bmlambda}{\bilambdad}
\safemath{\bmmu}{\bimud}
\safemath{\bmtheta}{\bithetad}
\safemath{\bmphi}{\biphid}
\safemath{\bmdelta}{\bideltad}

\safemath{\bA}{\mathbf{A}}
\safemath{\bB}{\mathbf{B}}
\safemath{\bC}{\mathbf{C}}
\safemath{\bD}{\mathbf{D}}
\safemath{\bE}{\mathbf{E}}
\safemath{\bF}{\mathbf{F}}
\safemath{\bG}{\mathbf{G}}
\safemath{\bH}{\mathbf{H}}
\safemath{\bI}{\mathbf{I}}
\safemath{\bJ}{\mathbf{J}}
\safemath{\bK}{\mathbf{K}}
\safemath{\bL}{\mathbf{L}}
\safemath{\bM}{\mathbf{M}}
\safemath{\bN}{\mathbf{N}}
\safemath{\bO}{\mathbf{O}}
\safemath{\bP}{\mathbf{P}}
\safemath{\bQ}{\mathbf{Q}}
\safemath{\bR}{\mathbf{R}}
\safemath{\bS}{\mathbf{S}}
\safemath{\bT}{\mathbf{T}}
\safemath{\bU}{\mathbf{U}}
\safemath{\bV}{\mathbf{V}}
\safemath{\bW}{\mathbf{W}}
\safemath{\bX}{\mathbf{X}}
\safemath{\bY}{\mathbf{Y}}
\safemath{\bZ}{\mathbf{Z}}

\safemath{\bZero}{\mathbf{0}}
\safemath{\bOne}{\mathbf{1}}
\safemath{\bDelta}{\mathbf{\Delta}}
\safemath{\bLambda}{\mathbf{\UpLambda}}
\safemath{\bPhi}{\mathbf{\Upphi}}
\safemath{\bSigma}{\mathbf{\Upsigma}}
\safemath{\bOmega}{\mathbf{\Upomega}}
\safemath{\bTheta}{\mathbf{\Uptheta}}

\bmdefine{\biAd}{A}
\bmdefine{\biBd}{B}
\bmdefine{\biCd}{C}
\bmdefine{\biDd}{D}
\bmdefine{\biEd}{E}
\bmdefine{\biFd}{F}
\bmdefine{\biGd}{G}
\bmdefine{\biHd}{H}
\bmdefine{\biId}{I}
\bmdefine{\biJd}{J}
\bmdefine{\biKd}{K}
\bmdefine{\biLd}{L}
\bmdefine{\biMd}{M}
\bmdefine{\biOd}{N}
\bmdefine{\biPd}{O}
\bmdefine{\biQd}{P}
\bmdefine{\biRd}{R}
\bmdefine{\biSd}{S}
\bmdefine{\biTd}{T}
\bmdefine{\biUd}{U}
\bmdefine{\biVd}{V}
\bmdefine{\biWd}{W}
\bmdefine{\biXd}{X}
\bmdefine{\biYd}{Y}
\bmdefine{\biZd}{Z}

\bmdefine{\biDelta}{\Delta}
\bmdefine{\biLambda}{\Lambda}
\bmdefine{\biPhi}{\Phi}
\bmdefine{\biSigma}{\Sigma}
\bmdefine{\biOmega}{\Omega}
\bmdefine{\biTheta}{\Theta}

\safemath{\bimA}{\biAd}
\safemath{\bimB}{\biBd}
\safemath{\bimC}{\biCd}
\safemath{\bimD}{\biDd}
\safemath{\bimE}{\biEd}
\safemath{\bimF}{\biFd}
\safemath{\bimG}{\biGd}
\safemath{\bimH}{\biHd}
\safemath{\bimI}{\biId}
\safemath{\bimJ}{\biJd}
\safemath{\bimK}{\biKd}
\safemath{\bimL}{\biLd}
\safemath{\bimM}{\biMd}
\safemath{\bimN}{\biNd}
\safemath{\bimO}{\biOd}
\safemath{\bimP}{\biPd}
\safemath{\bimQ}{\biQd}
\safemath{\bimR}{\biRd}
\safemath{\bimS}{\biSd}
\safemath{\bimT}{\biTd}
\safemath{\bimU}{\biUd}
\safemath{\bimV}{\biVd}
\safemath{\bimW}{\biWd}
\safemath{\bimX}{\biXd}
\safemath{\bimY}{\biYd}
\safemath{\bimZ}{\biZd}

\safemath{\bimDelta}{\biDelta}
\safemath{\bimLambda}{\biLambda}
\safemath{\bimPhi}{\biPhi}
\safemath{\bimSigma}{\biSigma}
\safemath{\bimOmega}{\biOmega}
\safemath{\bimTheta}{\biTheta}

\safemath{\setA}{\mathcal{A}}
\safemath{\setB}{\mathcal{B}}
\safemath{\setC}{\mathcal{C}}
\safemath{\setD}{\mathcal{D}}
\safemath{\setE}{\mathcal{E}}
\safemath{\setF}{\mathcal{F}}
\safemath{\setG}{\mathcal{G}}
\safemath{\setH}{\mathcal{H}}
\safemath{\setI}{\mathcal{I}}
\safemath{\setJ}{\mathcal{J}}
\safemath{\setK}{\mathcal{K}}
\safemath{\setL}{\mathcal{L}}
\safemath{\setM}{\mathcal{M}}
\safemath{\setN}{\mathcal{N}}
\safemath{\setO}{\mathcal{O}}
\safemath{\setP}{\mathcal{P}}
\safemath{\setQ}{\mathcal{Q}}
\safemath{\setR}{\mathcal{R}}
\safemath{\setS}{\mathcal{S}}
\safemath{\setT}{\mathcal{T}}
\safemath{\setU}{\mathcal{U}}
\safemath{\setV}{\mathcal{V}}
\safemath{\setW}{\mathcal{W}}
\safemath{\setX}{\mathcal{X}}
\safemath{\setY}{\mathcal{Y}}
\safemath{\setZ}{\mathcal{Z}}
\safemath{\emptySet}{\varnothing}

\safemath{\colA}{\mathscr{A}}
\safemath{\colB}{\mathscr{B}}
\safemath{\colC}{\mathscr{C}}
\safemath{\colD}{\mathscr{D}}
\safemath{\colE}{\mathscr{E}}
\safemath{\colF}{\mathscr{F}}
\safemath{\colG}{\mathscr{G}}
\safemath{\colH}{\mathscr{H}}
\safemath{\colI}{\mathscr{I}}
\safemath{\colJ}{\mathscr{J}}
\safemath{\colK}{\mathscr{K}}
\safemath{\colL}{\mathscr{L}}
\safemath{\colM}{\mathscr{M}}
\safemath{\colN}{\mathscr{N}}
\safemath{\colO}{\mathscr{O}}
\safemath{\colP}{\mathscr{P}}
\safemath{\colQ}{\mathscr{Q}}
\safemath{\colR}{\mathscr{R}}
\safemath{\colS}{\mathscr{S}}
\safemath{\colT}{\mathscr{T}}
\safemath{\colU}{\mathscr{U}}
\safemath{\colV}{\mathscr{V}}
\safemath{\colW}{\mathscr{W}}
\safemath{\colX}{\mathscr{X}}
\safemath{\colY}{\mathscr{Y}}
\safemath{\colZ}{\mathscr{Z}}

\safemath{\opA}{\mathbb{A}}
\safemath{\opB}{\mathbb{B}}
\safemath{\opC}{\mathbb{C}}
\safemath{\opD}{\mathbb{D}}
\safemath{\opE}{\mathbb{E}}
\safemath{\opF}{\mathbb{F}}
\safemath{\opG}{\mathbb{G}}
\safemath{\opH}{\mathbb{H}}
\safemath{\opI}{\mathbb{I}}
\safemath{\opJ}{\mathbb{J}}
\safemath{\opK}{\mathbb{K}}
\safemath{\opL}{\mathbb{L}}
\safemath{\opM}{\mathbb{M}}
\safemath{\opN}{\mathbb{N}}
\safemath{\opO}{\mathbb{O}}
\safemath{\opP}{\mathbb{P}}
\safemath{\opQ}{\mathbb{Q}}
\safemath{\opR}{\mathbb{R}}
\safemath{\opS}{\mathbb{S}}
\safemath{\opT}{\mathbb{T}}
\safemath{\opU}{\mathbb{U}}
\safemath{\opV}{\mathbb{V}}
\safemath{\opW}{\mathbb{W}}
\safemath{\opX}{\mathbb{X}}
\safemath{\opY}{\mathbb{Y}}
\safemath{\opZ}{\mathbb{Z}}
\safemath{\opZero}{\mathbb{O}}
\safemath{\identityop}{\opI}

\safemath{\veca}{\bma}
\safemath{\vecb}{\bmb}
\safemath{\vecc}{\bmc}
\safemath{\vecd}{\bmd}
\safemath{\vece}{\bme}
\safemath{\vecf}{\bmf}
\safemath{\vecg}{\bmg}
\safemath{\vech}{\bmh}
\safemath{\veci}{\bmi}
\safemath{\vecj}{\bmj}
\safemath{\veck}{\bmk}
\safemath{\vecl}{\bml}
\safemath{\vecm}{\bmm}
\safemath{\vecn}{\bmn}
\safemath{\veco}{\bmo}
\safemath{\vecp}{\bmp}
\safemath{\vecq}{\bmq}
\safemath{\vecr}{\bmr}
\safemath{\vecs}{\bms}
\safemath{\vect}{\bmt}
\safemath{\vecu}{\bmu}
\safemath{\vecv}{\bmv}
\safemath{\vecw}{\bmw}
\safemath{\vecx}{\bmx}
\safemath{\vecy}{\bmy}
\safemath{\vecz}{\bmz}

\safemath{\veczero}{\bmzero}
\safemath{\vecone}{\bmone}
\safemath{\vecxi}{\bmxi}
\safemath{\veclambda}{\bmlambda}
\safemath{\vecmu}{\bmmu}
\safemath{\vectheta}{\bmtheta}
\safemath{\vecphi}{\bmphi}
\safemath{\vecdelta}{\bmdelta}

\safemath{\matA}{\bA}
\safemath{\matB}{\bB}
\safemath{\matC}{\bC}
\safemath{\matD}{\bD}
\safemath{\matE}{\bE}
\safemath{\matF}{\bF}
\safemath{\matG}{\bG}
\safemath{\matH}{\bH}
\safemath{\matI}{\bI}
\safemath{\matJ}{\bJ}
\safemath{\matK}{\bK}
\safemath{\matL}{\bL}
\safemath{\matM}{\bM}
\safemath{\matN}{\bN}
\safemath{\matO}{\bO}
\safemath{\matP}{\bP}
\safemath{\matQ}{\bQ}
\safemath{\matR}{\bR}
\safemath{\matS}{\bS}
\safemath{\matT}{\bT}
\safemath{\matU}{\bU}
\safemath{\matV}{\bV}
\safemath{\matW}{\bW}
\safemath{\matX}{\bX}
\safemath{\matY}{\bY}
\safemath{\matZ}{\bZ}
\safemath{\matzero}{\bmzero}

\safemath{\matDelta}{\bDelta}
\safemath{\matLambda}{\bLambda}
\safemath{\matPhi}{\bPhi}
\safemath{\matSigma}{\bSigma}
\safemath{\matOmega}{\bOmega}
\safemath{\matTheta}{\bTheta}

\safemath{\matidentity}{\matI}
\safemath{\matone}{\matO}

\safemath{\rnda}{A}
\safemath{\rndb}{B}
\safemath{\rndc}{C}
\safemath{\rndd}{D}
\safemath{\rnde}{E}
\safemath{\rndf}{F}
\safemath{\rndg}{G}
\safemath{\rndh}{H}
\safemath{\rndi}{I}
\safemath{\rndj}{J}
\safemath{\rndk}{K}
\safemath{\rndl}{L}
\safemath{\rndm}{M}
\safemath{\rndn}{N}
\safemath{\rndo}{O}
\safemath{\rndp}{P}
\safemath{\rndq}{Q}
\safemath{\rndr}{R}
\safemath{\rnds}{S}
\safemath{\rndt}{T}
\safemath{\rndu}{U}
\safemath{\rndv}{V}
\safemath{\rndw}{W}
\safemath{\rndx}{X}
\safemath{\rndy}{Y}
\safemath{\rndz}{Z}

\safemath{\rveca}{\bimA}
\safemath{\rvecb}{\bimB}
\safemath{\rvecc}{\bimC}
\safemath{\rvecd}{\bimD}
\safemath{\rvece}{\bimE}
\safemath{\rvecf}{\bimF}
\safemath{\rvecg}{\bimG}
\safemath{\rvech}{\bimH}
\safemath{\rveci}{\bimI}
\safemath{\rvecj}{\bimJ}
\safemath{\rveck}{\bimK}
\safemath{\rvecl}{\bimL}
\safemath{\rvecm}{\bimM}
\safemath{\rvecn}{\bimN}
\safemath{\rveco}{\bomO}
\safemath{\rvecp}{\bimP}
\safemath{\rvecq}{\bimQ}
\safemath{\rvecr}{\bimR}
\safemath{\rvecs}{\bimS}
\safemath{\rvect}{\bimT}
\safemath{\rvecu}{\bimU}
\safemath{\rvecv}{\bimV}
\safemath{\rvecw}{\bimW}
\safemath{\rvecx}{\bimX}
\safemath{\rvecy}{\bimY}
\safemath{\rvecz}{\bimZ}

\safemath{\rvecxi}{\bmxi}
\safemath{\rveclambda}{\bmlambda}
\safemath{\rvecmu}{\bmmu}
\safemath{\rvectheta}{\bmtheta}
\safemath{\rvecphi}{\bmphi}

\safemath{\rmatA}{\bimA}
\safemath{\rmatB}{\bimB}
\safemath{\rmatC}{\bimC}
\safemath{\rmatD}{\bimD}
\safemath{\rmatE}{\bimE}
\safemath{\rmatF}{\bimF}
\safemath{\rmatG}{\bimG}
\safemath{\rmatH}{\bimH}
\safemath{\rmatI}{\bimI}
\safemath{\rmatJ}{\bimJ}
\safemath{\rmatK}{\bimK}
\safemath{\rmatL}{\bimL}
\safemath{\rmatM}{\bimM}
\safemath{\rmatN}{\bimN}
\safemath{\rmatO}{\bimO}
\safemath{\rmatP}{\bimP}
\safemath{\rmatQ}{\bimQ}
\safemath{\rmatR}{\bimR}
\safemath{\rmatS}{\bimS}
\safemath{\rmatT}{\bimT}
\safemath{\rmatU}{\bimU}
\safemath{\rmatV}{\bimV}
\safemath{\rmatW}{\bimW}
\safemath{\rmatX}{\bimX}
\safemath{\rmatY}{\bimY}
\safemath{\rmatZ}{\bimZ}

\safemath{\rmatDelta}{\bimDelta}
\safemath{\rmatLambda}{\bimLambda}
\safemath{\rmatPhi}{\bimPhi}
\safemath{\rmatSigma}{\bimSigma}
\safemath{\rmatOmega}{\bimOmega}
\safemath{\rmatTheta}{\bimTheta}

\usepackage{amssymb}
\usepackage{amsfonts}
\usepackage{mathrsfs}
\usepackage{xspace}
\usepackage{bm}
\usepackage{fancyref}
\usepackage{textcomp}

\usepackage{multirow}
\usepackage{stmaryrd}

\newenvironment{textbmatrix}{	\setlength{\arraycolsep}{2.5pt}%
	\big[\begin{matrix}}{\end{matrix}\big]%
	\raisebox{0.08ex}{\vphantom{M}}}

\def\be{\begin{equation}}
	\def\ee{\end{equation}}
\def\een{\nonumber \end{equation}}
\def\mat{\begin{bmatrix}}
\def\emat{\end{bmatrix}}
\def\btm{\begin{textbmatrix}}
\def\etm{\end{textbmatrix}}

\def\ba#1\ea{\begin{align}#1\end{align}}
\def\bas#1\eas{\begin{align*}#1\end{align*}}
\def\bs#1\es{\begin{split}#1\end{split}}
\def\bg#1\eg{\begin{gather}#1\end{gather}}
\def\bml#1\eml{\begin{multline}#1\end{multline}}
\def\bi#1\ei{\begin{itemize}#1\end{itemize}}

\safemath{\dirac}{\delta}					
\safemath{\krond}{\dirac}					

\safemath{\upto}{\uparrow}
\safemath{\downto}{\downarrow}
\safemath{\iu}{j}							
\safemath{\ev}{\lambda}						
\safemath{\hilseqspace}{l^{2}}				
\newcommand{\banachfunspace}[1]{\setL^{#1}}	
\safemath{\hilfunspace}{\banachfunspace{2}}	

\safemath{\SNR}{\textit{SNR}} 				
\safemath{\PAR}{\textit{PAR}} 				
\safemath{\No}{N_0}							
\safemath{\Es}{E_s}							
\safemath{\Eb}{E_b}							
\safemath{\EbNo}{\frac{\Eb}{\No}}
\safemath{\EsNo}{\frac{\Es}{\No}}

\DeclareMathOperator{\CHop}{\ensuremath{\opH}} 
\safemath{\tvir}{\rndh_{\CHop}}				
\safemath{\tvtf}{\rndl_{\CHop}}				
\safemath{\spf}{\rnds_{\CHop}}				
\safemath{\bff}{H_{\CHop}}					

\safemath{\ircf}{r_{h}}						
\safemath{\tftvcf}{r_{s}}					
\safemath{\tfcf}{r_{l}}						
\safemath{\bfcf}{r_{H}}						

\safemath{\tcorr}{c_h}						
\safemath{\scf}{c_{s}}						
\safemath{\tfcorr}{c_{l}}					
\safemath{\fcorr}{c_{H}}						

\safemath{\mi}{I}							
\safemath{\capacity}{C}						

\safemath{\normal}{\mathcal{N}}			
\safemath{\jpg}{\mathcal{CN}}			
\safemath{\mchain}{\leftrightarrow}		

\safemath{\dB}{\,\mathrm{dB}}
\safemath{\dBm}{\,\mathrm{dBm}}
\safemath{\Hz}{\,\mathrm{Hz}}
\safemath{\kHz}{\,\mathrm{kHz}}
\safemath{\MHz}{\,\mathrm{MHz}}
\safemath{\GHz}{\,\mathrm{GHz}}
\safemath{\s}{\,\mathrm{s}}
\safemath{\ms}{\,\mathrm{ms}}
\safemath{\mus}{\,\mathrm{\text{\textmu}s}}
\safemath{\ns}{\,\mathrm{ns}}
\safemath{\ps}{\,\mathrm{ps}}
\safemath{\meter}{\,\mathrm{m}}
\safemath{\mm}{\,\mathrm{mm}}
\safemath{\cm}{\,\mathrm{cm}}
\safemath{\m}{\,\mathrm{m}}
\safemath{\W}{\,\mathrm{W}}
\safemath{\mW}{\, \mathrm{mW}}
\safemath{\J}{\,\mathrm{J}}
\safemath{\K}{\,\mathrm{K}}
\safemath{\bit}{\,\mathrm{bit}}
\safemath{\nat}{\,\mathrm{nat}}

\safemath{\define}{\triangleq}			

\safemath{\equivalent}{\sim}
\safemath{\distas}{\sim}					
\safemath{\sdiff}{\Delta}				

\safemath{\reals}{\mathbb{R}}
\safemath{\positivereals}{\reals_{+}}
\safemath{\integers}{\mathbb{Z}}
\safemath{\posint}{\integers_{+}}
\safemath{\naturals}{\mathbb{N}}
\safemath{\posnaturals}{\naturals_{+}}
\safemath{\complexset}{\mathbb{C}}
\safemath{\rationals}{\mathbb{Q}}

\newcommand*{\fancyrefapplabelprefix}{app}		
\newcommand*{\fancyrefthmlabelprefix}{thm}		
\newcommand*{\fancyreflemlabelprefix}{lem}		
\newcommand*{\fancyrefcorlabelprefix}{cor}		
\newcommand*{\fancyrefdeflabelprefix}{def}		
\newcommand*{\fancyrefproplabelprefix}{prop}		
\newcommand*{\fancyrefexmpllabelprefix}{exmpl}
\newcommand*{\fancyrefalglabelprefix}{alg}		
\newcommand*{\fancyreftbllabelprefix}{tbl}		

\frefformat{vario}{\fancyrefseclabelprefix}{Sec.~#1}
\frefformat{vario}{\fancyrefthmlabelprefix}{Theorem~#1}
\frefformat{vario}{\fancyreftbllabelprefix}{Tbl.~#1}
\frefformat{vario}{\fancyreflemlabelprefix}{Lemma~#1}
\frefformat{vario}{\fancyrefcorlabelprefix}{Corollary~#1}
\frefformat{vario}{\fancyrefdeflabelprefix}{Definition~#1}
\frefformat{vario}{\fancyreffiglabelprefix}{Fig.~#1}
\frefformat{vario}{\fancyrefapplabelprefix}{Appendix~#1}
\frefformat{vario}{\fancyrefeqlabelprefix}{(#1)}
\frefformat{vario}{\fancyrefproplabelprefix}{Proposition~#1}
\frefformat{vario}{\fancyrefexmpllabelprefix}{Example~#1}
\frefformat{vario}{\fancyrefalglabelprefix}{Algorithm~#1}

\safemath{\dictab}{[\,\dicta\,\,\dictb\,]}

\safemath{\ysig}{\bmy}
\safemath{\ysighat}{\hat{\ysig}}
\safemath{\ysigdim}{M}
\safemath{\xsig}{\bmx}
\safemath{\xsigdim}{N}
\safemath{\nx}{n_x}
\safemath{\zsig}{\bmz}
\safemath{\zsigdim}{\ysigdim}
\safemath{\rsig}{\bmr}
\safemath{\Adict}{\bA}
\safemath{\Adicttilde}{\widetilde{\Adict}}
\safemath{\Adictdim}{\outputdim\times\xsigdim}
\safemath{\avec}{\bma}
\safemath{\avectilde}{\tilde{\avec}}
\safemath{\Bdict}{\bB}
\safemath{\Bdicttilde}{\widetilde{\Bdict}}
\safemath{\Cdict}{\bC}
\safemath{\cvec}{\bmc}
\safemath{\Ddict}{\bD}
\safemath{\Ddictdim}{\ysigdim\times\xsigdim}
\safemath{\dvec}{\bmd}
\safemath{\Ddicttilde}{\widetilde{\bD}}
\safemath{\Bonb}{\bB}
\safemath{\bvec}{\bmb}
\safemath{\Bonbdim}{\ysigdim\times\ysigdim}
\safemath{\noise}{\bmn}
\safemath{\noisedim}{\ysigim}
\safemath{\err}{\bme}
\safemath{\errdim}{\ysigdim}
\safemath{\errset}{\setE}
\safemath{\nerr}{n_e}
\safemath{\delop}{\bP_\errset}
\safemath{\delopc}{\bP_{{\errset}^c}}

\safemath{\cplxi}{\imath}
\safemath{\cplxj}{\jmath}

\safemath{\dict}{\matD}
\safemath{\inputdim}{N}		
\safemath{\outputdim}{M}		
\safemath{\sparsity}{S}	
\safemath{\inputdimA}{{N_a}}	
\safemath{\inputdimB}{{N_b}}	
\safemath{\elemA}{{n_a}}	
\safemath{\elemB}{{n_b}}	
\safemath{\resA}{\matR_a}	
\safemath{\resB}{\matR_b}	
\safemath{\subD}{\matS} 
\safemath{\subA}{\matS_a} 
\safemath{\subB}{\matS_b} 
\safemath{\dicta}{\matA} 	
\safemath{\dictb}{\matB} 	
\safemath{\hollowS}{H}
\safemath{\hollowA}{H_a}
\safemath{\hollowB}{H_b}
\safemath{\cross}{Z}
\safemath{\coh}{\mu_d}			
\safemath{\coha}{\mu_a}			
\safemath{\cohb}{\mu_b}			
\safemath{\mubs}{\nu}	
\safemath{\cohm}{\mu_m} 
\safemath{\dictset}{\setD}	
\safemath{\dictsetp}{\dictset(\coh,\coha,\cohb)}	
\safemath{\dictsetgen}{\dictset_\text{gen}}
\safemath{\dictsetgenp}{\dictsetgen(\coh)}
\safemath{\dictsetonb}{\dictset_\text{onb}}
\safemath{\dictsetonbp}{\dictsetonb(\coh)}

\safemath{\leftside}{U}
\safemath{\rightsideA}{R_a}
\safemath{\rightsideB}{R_b}

\safemath{\indexS}{\setI_S} 

\safemath{\na}{n_a}			
\safemath{\nb}{n_b}			
\safemath{\coeffa}{p_i}	
\safemath{\coeffb}{q_j}	
\safemath{\seta}{\setP}		
\safemath{\setb}{\setQ}     
\safemath{\setw}{\setW}	
\safemath{\setz}{\setZ}	
\safemath{\cola}{\veca}		
\safemath{\colb}{\vecb}		
\safemath{\cold}{\vecd}		
\safemath{\inputvec}{\vecx} 	
\safemath{\error}{\vece}	
\safemath{\noiseout}{\vecz} 	
\safemath{\inputvecel}{x}
\safemath{\inputveca}{\vecx_a}
\safemath{\inputvecb}{\vecx_b}
\safemath{\outputvec}{\vecy}	
\safemath{\lambdamin}{\lambda_{\mathrm{min}}}

\safemath{\elltwo}{\ell_2}
\safemath{\ellone}{\ell_1}
\safemath{\ellzero}{\ell_0}
\safemath{\ellinf}{\ell_\infty}
\safemath{\ellinftilde}{\ell_{\widetilde\infty}}
\safemath{\licard}{Z(\coh,\coha,\cohb)}
\safemath{\xsol}{\hat{x}}
\safemath{\xbord}{x_b}		
\safemath{\xstat}{x_s}		
\safemath{\xstatLone}{\tilde{x}_s}
\safemath{\order}{\mathcal{O}} 
\safemath{\scales}{\Theta} 
\safemath{\ones}{\mathbf{1}} 
\safemath{\zeroes}{\mathbf{0}} 
\safemath{\thlone}{\kappa(\coh,\cohb)} 
\safemath{\constoneA}{\delta} 
\safemath{\constoneB}{\epsilon} 
\safemath{\nlarge}{L}				   
\safemath{\sumlarge}{S_\nlarge}
\safemath{\maxlarger}{P_\nlarge}	   
\safemath{\Pzero}{\textrm{P0}}	
\safemath{\Pone}{\textrm{P1}}
\safemath{\vecfir}{\vecw}			 
\safemath{\vecsec}{\vecz}
\safemath{\elvecfir}{w}              
\safemath{\elvecsec}{z}				 
\safemath{\nlargefir}{n}
\safemath{\normout}{\gamma}
\safemath{\auxfun}{h}
\safemath{\supp}{\textrm{supp}}

\safemath{\indexa}{\ell}
\safemath{\indexb}{r}
\safemath{\indexc}{i}
\safemath{\indexd}{j}

\safemath{\project}{P}

\allowdisplaybreaks

\newcommand{\revision}[1]{#1}

\safemath{\LAMA}{\textrm{LAMA}}
\safemath{\MRT}{\textrm{MRT}}
\safemath{\betamax}{\beta^\text{max}_\setO}
\safemath{\betamaxno}{\beta^\text{max}}
\safemath{\betamin}{\beta^\text{min}_\setO}
\safemath{\betaminno}{\beta^\text{min}}

\safemath{\Nomin}{\No^\textnormal{min}(\beta)}
\safemath{\Nominnobeta}{\No^\text{min}}
\safemath{\Nomax}{\No^\textnormal{max}(\beta)}
\safemath{\Nomaxnobeta}{\No^\textnormal{max}}
\safemath{\EX}{E_\textnormal{x}}
\safemath{\EXP}{\EX^\textnormal{p}}
\safemath{\Eo}{E_0}

\safemath{\tmax}{{t_\textnormal{max}}}
\safemath{\MAP}{\textrm{MAP}}
\safemath{\IO}{\textrm{IO}}
\safemath{\JO}{\textrm{JO}}
\safemath{\Nopost}{N_{0}^\textnormal{post}}
\safemath{\MT}{U}
\safemath{\MR}{B}
\safemath{\Tran}{\textnormal{T}}
\safemath{\Herm}{\textnormal{H}}
\safemath{\row}{\textnormal{r}}
\safemath{\col}{\textnormal{c}}

\safemath{\NT}{N_\textnormal{T}}
\safemath{\DSNR}{\delta \textnormal{SNR}}
\safemath{\betaMOR}{\beta^{\star}}

\begin{document}

\title{A 46\,Gbps 12\,pJ/b Sparsity-Adaptive Beamspace Equalizer for mmWave Massive MIMO in 22FDX}

\author{Seyed Hadi Mirfarshbafan and Christoph Studer%
\thanks{S.~H.~Mirfarshbafan and C.~Studer are with the Department of Information Technology and Electrical Engineering, ETH Z\"urich, Switzerland (email: mirfarshbafan@iis.ee.ethz.ch and studer@ethz.ch).}
\thanks{The authors thank GlobalFoundries for providing silicon fabrication through the 22FDX University Program.}
}

\maketitle
	
\begin{abstract}
	We present a GlobalFoundries 22FDX FD-SOI application-specific integrated circuit (ASIC) of a beamspace equalizer for millimeter-wave (mmWave) massive multiple-input multiple-output (MIMO) systems. The ASIC implements a recently-proposed power-saving technique called sparsity-adaptive equalization (SPADE). SPADE exploits the inherent sparsity of mmWave channels in the beamspace domain to reduce the dynamic power of matrix-vector products by skipping multiplications for which the magnitude of both operands are below pre-defined thresholds. Simulations with realistic mmWave channels show that SPADE incurs less than 0.7\,dB SNR degradation at 1\% target bit error rate compared to antenna-domain equalization. ASIC measurement results demonstrate an equalization throughput of 46\,Gbps and show that SPADE offers up to 38\% power savings compared to antenna-domain equalization. A comparison with state-of-the-art \revision{massive MIMO equalizer designs reveals} that our ASIC achieves superior normalized energy efficiency.  
\end{abstract}


\section{Introduction}

Fifth generation (5G) and \revision{beyond-5G} wireless communication systems take advantage of large contiguous portions of the available spectrum at millimeter-wave (mmWave) frequencies to enable wideband communication~\cite{3gpp22}. 
Corresponding basestations (BSs) rely on massive multiple-input multiple-output (MIMO)~\cite{ericsson22}, which (i) mitigates the high path loss at mmWave frequencies~\cite{rappaport15a} and (ii) enables multi-user (MU) communication by means of spatial multiplexing.
\revision{Wideband communication requires high baseband sampling rates and massive MU-MIMO generates high-dimensional data---together, they   significantly increase hardware complexity. In this paper, we present a hardware implementation of a technique that reduces the power consumption of data detection.}

\subsubsection{Beamspace Processing}

A promising approach to reducing complexity of data detection in \revision{all-digital} mmWave massive MU-MIMO systems is to exploit the inherent sparsity of mmWave channels\mbox{\cite{schniter14a, rappaport15a}} in the so-called \emph{beamspace}.
Converting a system from antenna-domain into beamspace is achieved by applying a spatial discrete Fourier transform (DFT) \revision{to} the signals received at a uniform linear antenna array~\cite{Abdelghany19, Mahdavi20, SeyedHadi20b, mirfarshbafan21b, goluntas21, mirfarshbafan19a}.
Uplink data detection in beamspace, with the goal of reducing implementation complexity, has been studied recently for mmWave massive MU-MIMO systems, mainly in the context of linear data detectors, as nonlinear methods typically incur higher complexity. Linear data detection consists of two phases: (i) \emph{preprocessing}, where an equalization matrix is computed based on a channel-matrix estimate and (ii) \emph{equalization}, where the equalization matrix is multiplied to the received vectors to obtain estimates of the transmitted data symbols. While  preprocessing is performed only once per coherence interval, equalization must be performed for each received vector, hence, at much higher rates than preprocessing. \revision{In this paper, we focus on reducing the complexity of equalization and assume that preprocessing is performed externally.}

Existing beamspace data detectors reduce equalization complexity by designing \emph{sparse}  equalization matrices with specific sparsity patterns, thereby reducing the number of multiplications required  for equalization. Such \revision{sparsity-exploiting} beamspace data detectors, however, either incur a notable performance degradation compared to conventional antenna-domain linear minimum mean squared error (LMMSE) \revision{equalization,} e.g.,~\cite{Abdelghany19, Mahdavi20}, or require preprocessing algorithms with extremely high computational complexity~\cite{SeyedHadi20b}.

In \cite{mirfarshbafan21b}, a different approach to reduce complexity by exploiting beamspace sparsity was proposed. The method is referred to as \emph{sparsity-adaptive equalization} (SPADE) and leverages the fact that the LMMSE equalization matrix is already \emph{approximately sparse} in beamspace and avoids computing a sparse equalization matrix with a specific sparsity pattern.
To reduce equalization complexity, SPADE uses two pre-computed thresholds to skip multiplications whenever the absolute value of both operands are below these thresholds. \revision{As shown in~\cite{mirfarshbafan21b}, SPADE significantly reduces the number of required multiplications, while exhibiting comparable  performance to state-of-the-art linear beamspace data  detectors~\cite{Abdelghany19, Mahdavi20,SeyedHadi20b}.}

\subsubsection{Contributions}
We present the first application specific integrated circuit (ASIC) capable of performing SPADE-based beamspace equalization as well as antenna-domain equalization for a massive MU-MIMO system with $64$ BS antennas and up to $16$ single-antenna user equipments (UEs). 
In addition, we demonstrate real-world power savings achieved by SPADE-based beamspace equalization over conventional, antenna-domain equalization through extensive ASIC measurements.

\subsubsection{Notation} \label{sec:notation}
Boldface lowercase and uppercase letters represent column vectors and matrices, respectively. For a matrix $\bA$, the transpose is $\bA^\Tran$ and Hermitian transpose $\bA^\Herm$. The $m$th column of $\bA$ is $\bma_m = [\bA]_m$, and the entry on the $m$th row and $n$th column  is $A_{m,n} = [\bA]_{m,n}$.
For a vector $\bma$, the $k$th entry is $a_k = [\bma]_k$, and the real and imaginary parts are~$\bma^R$ and $\bma^I$, respectively. 
The $\ell_\infty$- and $\ell_{\widetilde\infty}$-norm is $\|\bma\|_\infty \define \max_{k} |a_k|$ and $\|\bma\|_{\widetilde\infty} \define \max\{\|\bma^R\|_\infty,\|\bma^{I}\|_\infty\}$, respectively \cite{seethaler10a}.
Bars over variables indicate antenna-domain quantities.
Expectation with respect to a random vector~$\bma$ is denoted by~$\mathbb{E}_{\bma}[\cdot]$.


\section{Prerequisites} 
\label{sec:background}

\subsection{Antenna-Domain System Model}
\label{sec:beamspace}

\begin{figure}[tp]
\centering
\includegraphics[width=0.9\columnwidth]{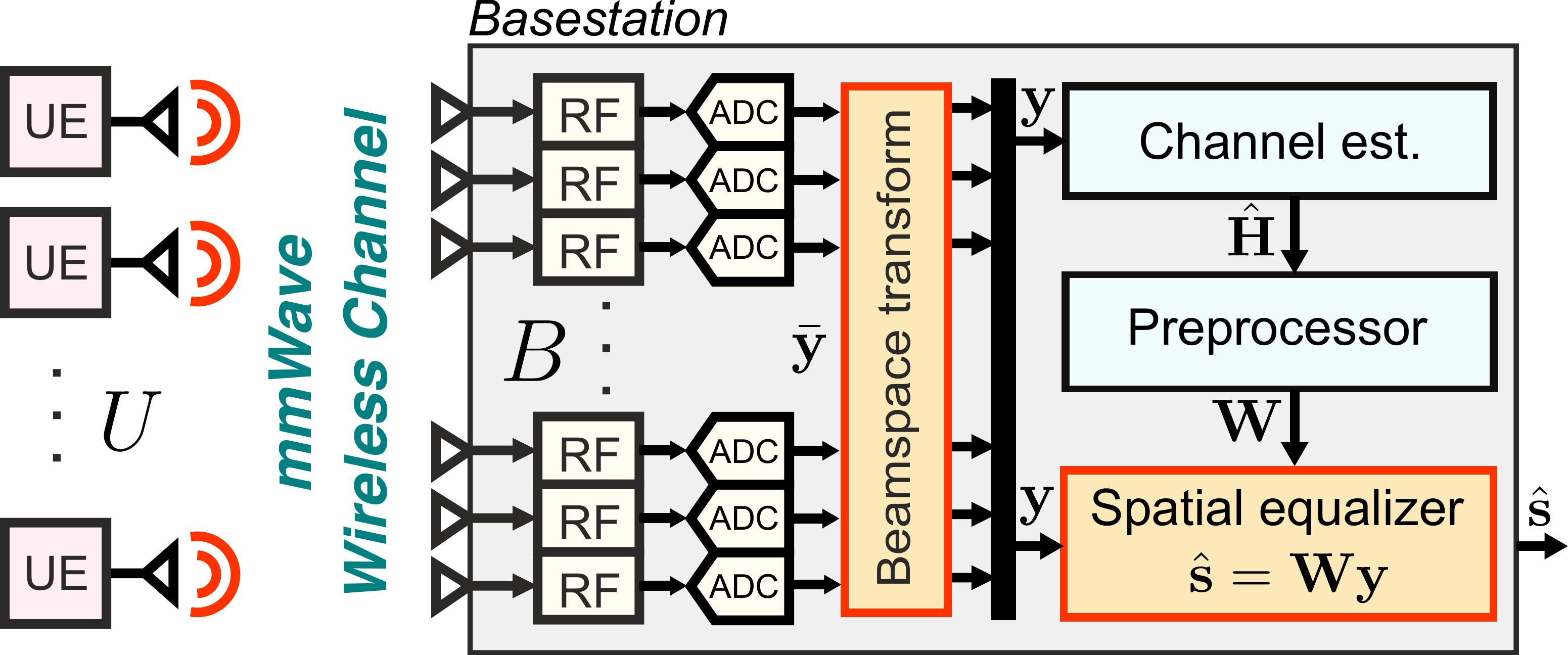}
\vspace{-0.1cm}
\caption{Beamspace processing in mmWave massive MU-MIMO systems}
\label{fig:system_overview}
\end{figure}

We consider the uplink of an all-digital mmWave massive MU-MIMO system as depicted in \fref{fig:system_overview}. Here, $U$ single-antenna UEs transmit data to a basestation that is equipped with a $B$-antenna uniform linear  array.
The \emph{antenna-domain} received vector at the BS is given by\footnote{Although this model only holds for frequency-flat channels, an extension of this paper's results to frequency-selective channels is straightforward if using orthogonal frequency-division multiplexing (OFDM).}
$\bar{\bmy} = \bar{\bH} \bms + \bar{\bmn}$,
where $\bar{\bmy}\in\complexset^B$ is the antenna-domain receive vector, \mbox{$\bar{\bH} \in \complexset^{B\times U}$} is the antenna-domain MIMO channel matrix, $\bms\in\complexset^U$ contains the transmit symbols of the $U$ UEs (taken from a constellation set), and $\bar{\bmn}\in\complexset^B$ models noise with i.i.d.\ circularly symmetric complex Gaussian entries with variance $N_0$. 
The transmit symbols are assumed to satisfy the power constraint $\mathbb{E}_{\bms}[\bms \bms^\Herm]=\Es \bI_U$.

\subsection{Beamspace System Model}
Using the well-known planar-wave approximation\footnote{\revision{We use the planar-wave approximation for mathematical analysis---our simulations, however, use channel vectors generated with spherical waves.}} \cite{tse05a}, we model the wireless channel between a UE and the BS as
\begin{align}
	\label{eq:planarwave}
	\bar{\bmh} = \textstyle \sum_{\ell=0}^{L-1} \alpha_\ell \> \bar{\bma}(\phi_\ell),
\end{align}
where $L$ refers to the number of propagation paths, $\alpha_\ell\in\opC$ is the complex-valued channel gain of the $\ell$th path, and
\begin{align} \label{eq:complexsinusoids}
	\bar{\bma}(\phi_\ell) = \big[1, e^{j\phi_\ell},e^{j2\phi_\ell},\dots, e^{j(B-1)\phi_\ell} \big]^\Tran,
\end{align}
where the spatial frequency $\phi_\ell$ is determined by the $\ell$th path's angle-of-arrival at the BS antenna array. Due to the predominantly directional nature of wave propagation at mmWave frequencies~\cite{miao23}, $L$ is typically much smaller than $B$ in massive MIMO systems, meaning that the channel vector of each UE is a superposition of only a few complex sinusoids. Hence, the DFT of $\bar{\bmh}$ in \fref{eq:planarwave} results in an approximately sparse vector \cite{mirfarshbafan19a}, meaning that most of its entries are close to zero, while only few entries have large magnitude.
By applying such a spatial DFT to the antenna-domain received vector $\bar{\bmy}$, we arrive at the following \emph{beamspace} system model:
\begin{align} \label{eq:BD_model}
	\bmy = \bF\bar{\bmy} = \bF\bar{\bH}\bms + \bF\bar{\bmn} 	= \bH \bms + \bmn.
\end{align} 
Here, $\bmy\in\complexset^B$ is the \emph{beamspace} receive vector,  $\bF\in\complexset^{B\times B}$ is the unitary DFT matrix, $\bH = \bF\bar{\bH}$ is the beamspace MIMO channel matrix, and $\bmn = \bF\bar{\bmn}$ is the beamspace-equivalent noise vector, which has the same statistics as $\bar{\bmn}$ as $\bF$ is unitary.
Therefore, beamspace system model in \fref{eq:BD_model} is statistically equivalent to the antenna-domain system model, and data detection using both models gives exactly the same result.

\subsection{SParsity-ADaptive Equalization (SPADE)} \label{sec:spade} 

We now summarize SPADE~\cite{mirfarshbafan21b}, 
which is the main ingredient of our ASIC.
First, we note that due to the approximate sparsity of the beamspace channel matrix $\bH$, the beamspace LMMSE equalization matrix $\bV = (\bH^\Herm \bH + \frac{\No}{\Es} \bI)^{-1}\bH^\Herm$ also exhibits approximate sparsity. Second, since the beamspace receive vector $\bmy$ is a linear combination of a few sparse vectors, it is also approximately sparse.
SPADE exploits this approximate sparsity in both~$\bV$ and $\bmy$ in order to reduce the number of effective multiplications during equalization  $\hat{\bms} = \bV \bmy$. 

Consider the $u$th inner product $\hat{s}_u  = \sum_{b=1}^B V_{u,b} y_b$. 
Due to approximate sparsity of $\bmy$ and the rows of $\bV$, many of the operands~$V_{u,b}$ and~$y_b$ have small magnitude.
Since  the number of BS antennas $B$ is large in massive MIMO, each such inner product is a sum of a large number of products. Consequently, one can skip multiplications where both~$|V_{u,b}|$ and~$|y_b|$ are small, without a notable perturbation on the inner-product's result. 
Noting that each complex-valued multiplication consists of four real-valued multiplications, instead of skipping an entire complex-valued multiplication, one can individually turn on or off the real-valued multiplications based on the absolute values of their operands.
SPADE uses two thresholds $\tau_y$ and~$\tau_w$ to skip real-valued multiplications for which the absolute value of both operands are below the respective thresholds. These thresholds trade computational complexity for accuracy of the inner product and are determined offline based on simulations with the goal of minimizing (i) the approximation error and (ii) the \emph{multiplier activity rate}, which is the average number of executed real-valued multiplications divided by the total $4BU$ real-valued multiplications involved in $\bV \bmy$.

Note that the rows of $\bV$ may have vastly different dynamic ranges, calling for a separate threshold $\tau_w$ for each row. In order to  use the same threshold $\tau_w$ for all of the inner products, the rows of $\bV$ are scaled to obtain $\bW = \mathrm{diag}(\boldsymbol\alpha)\bV$, where $\alpha_u=1/(\|[\bV^\Tran]_u\|_{\widetilde\infty}+\varepsilon)$, and $\varepsilon>0$ is a small constant that ensures that $\|[\bW^\Tran]_u\|_{\widetilde\infty}$ is just below one. \revision{The final symbol estimates are then obtained as $\hat{\bms} = \mathrm{diag}(\boldsymbol\alpha)^{-1} \bW \bmy$. This row-wise scaling has the additional benefit of reducing the overall dynamic range of the entries in~$\bV$, thereby reducing the minimum required bitwidth of entries in $\bW$ in its fixed-point representation}. 

\subsection{Error-Rate Performance Simulations} \label{sec:sim}
\begin{figure}[t]
	\centering
	\vspace{-0.2cm}
	\subfigure[$64\times 16$, non-LoS]
	{\includegraphics[width=0.48\columnwidth]{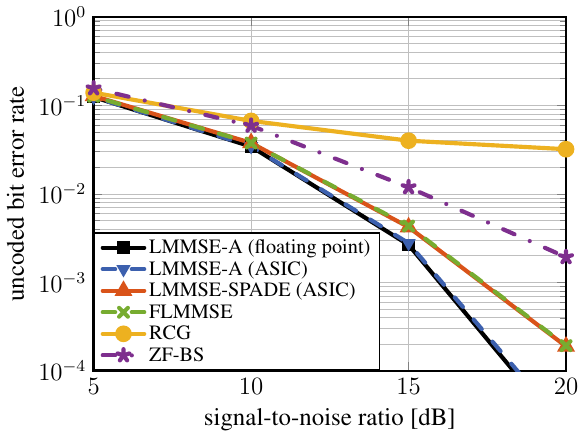}\label{fig:NLOS64x16}}
	\hfill
	\subfigure[$64\times 16$, LoS]
	{\includegraphics[width=0.48\columnwidth]{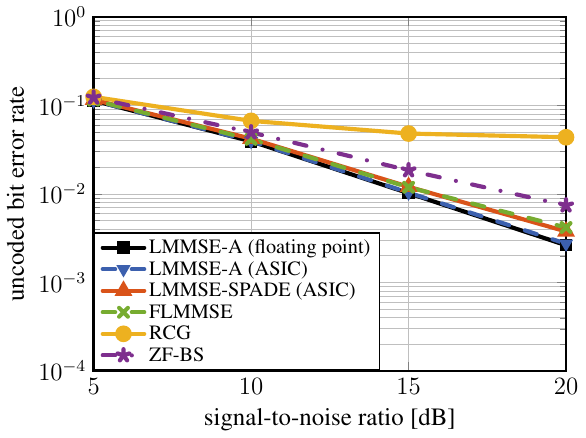}\label{fig:LOS64x16}}
	\subfigure[$64\times 8$, non-LoS]
	{\includegraphics[width=0.48\columnwidth]{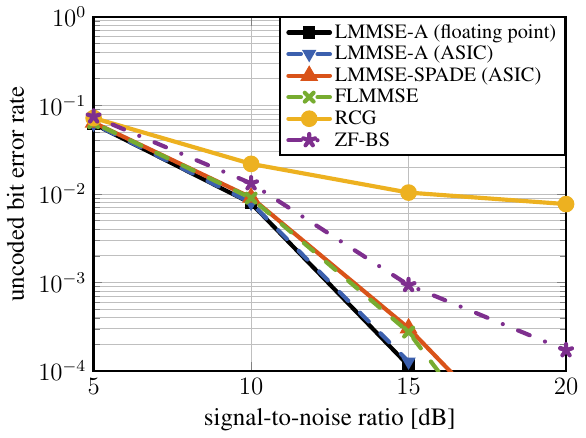}\label{fig:NLOS64x8}}
	\hfill
	\subfigure[$64\times 8$, LoS]
	{\includegraphics[width=0.48\columnwidth]{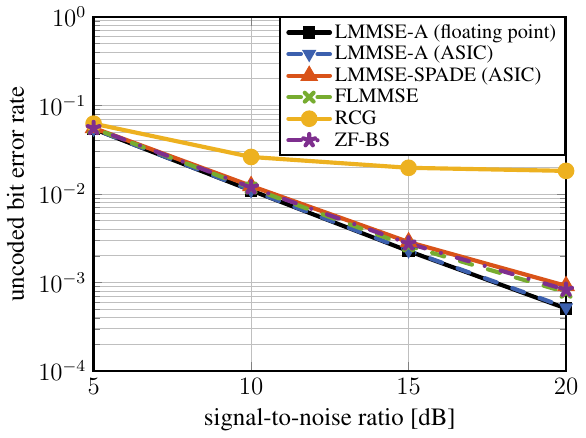}\label{fig:LOS64x8}}
	
	\vspace{-0.1cm}
	\caption{Uncoded BER simulation results of LMMSE-A with floating-point operations, fixed-point LMMSE-A and LMMSE-SPADE as implemented in our ASIC, as well as existing baseline algorithms.}
	\label{fig:BER}
\end{figure}

In order to evaluate the impact of SPADE's approximate inner-product computations on the system performance, we simulate \revision{the} uncoded bit error-rate (BER) for beamspace LMMSE employing SPADE (referred to as `LMMSE-SPADE').
\fref{fig:BER} shows the results for LMMSE-SPADE with $16$ or~$8$ UEs transmitting $16$-QAM symbols to a $64$-antenna BS over line-of-sight (LoS)\footnote{\revision{The generated LoS channel vectors also include reflective paths.}} and non-LoS channels generated with the QuadRiGa mmMAGIC simulator~\cite{QuaDRiGa_tech_rpt}. We also show the BER of the antenna-domain mode of our ASIC, labeled ``\mbox{LMMSE-A} (ASIC),'' as well as a floating-point reference.
We observe that LMMSE-SPADE incurs less than $0.4$\,dB and $0.7$\,dB SNR loss at $1$\% BER for $U=8$ and $U=16$, respectively, compared to LMMSE-A. The threshold pairs corresponding to these simulations and the ASIC measurements are detailed in \fref{sec:threshols}.
Furthermore, we compare the BER to three state-of-the-art massive MU-MIMO \revision{data detection algorithms}: FLMMSE~\cite{sirpac21}, recursive conjugate gradient (RCG)~\cite{liu20}, and zero-forcing with beam selection (ZF-BS)~\cite{Mahdavi20}. While FLMMSE exhibits similar performance to LMMSE-SPADE, RCG suffers a performance loss in highly correlated mmWave channels, and ZF-BS suffers a notable performance loss in non-LoS channels.

\subsection{How to Select the SPADE Thresholds?}
\label{sec:threshols}
\begin{figure}[t]
	\vspace{-0.2cm}
	\centering
	\subfigure[$64 \times 16$, non-LoS]
	{\includegraphics[width=0.47\columnwidth]{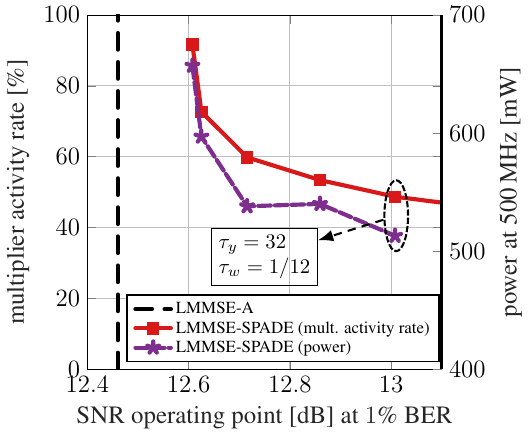}\label{fig:threshods_NLOS}}
	\hfill
	\subfigure[$64 \times 16$, LoS]
	{\includegraphics[width=0.47\columnwidth]{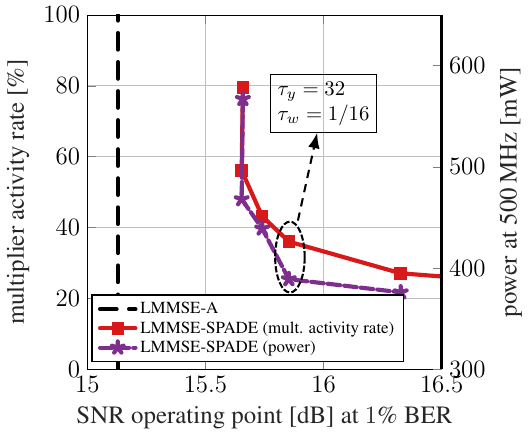}\label{fig:threshods_LOS}}
	\vspace{-0.1cm}
	\caption{Multiplier activity rate vs.~SNR operating point at $1\%$ BER in a $64 \times 16$ system for several threshold pairs for (a) non-LoS and (b) LoS channels. The thresholds for our ASIC measurements were determined offline.}
	\vspace{-0.0cm}
	\label{fig:thresholds}
\end{figure}

\revision{The SPADE thresholds $\tau_y$ and~$\tau_w$ determine the inner-product accuracy and the multiplier activity rate. To simplify  implementation and to eliminate the need for determining these thresholds on-the-fly for each considered system dimension and channel condition (LoS or non-LoS), we found a fixed pair of thresholds that achieves a desirable trade-off between BER and power savings. To this end, we performed BER simulations offline for a range of threshold pairs, and for each pair, we computed the average multiplier activity rate and the \emph{SNR operating point} (i.e., the minimum SNR that achieves $1\%$ BER).
\fref{fig:thresholds} provides the results for a system with $B=64$ BS antennas and $U=16$ UEs, in which we also show the power consumption corresponding to each threshold pair, extracted from stimuli-based power simulations at $500$\,MHz (on the right y-axis), as well as the chosen threshold pair. Evidently, the multiplier activity well-predicts the power consumption.
}

\section{VLSI Architecture} 
\label{sec:vlsi}

\begin{figure}[tp]
\vspace{-0.18cm}
\centering
\includegraphics[width=0.9\columnwidth]{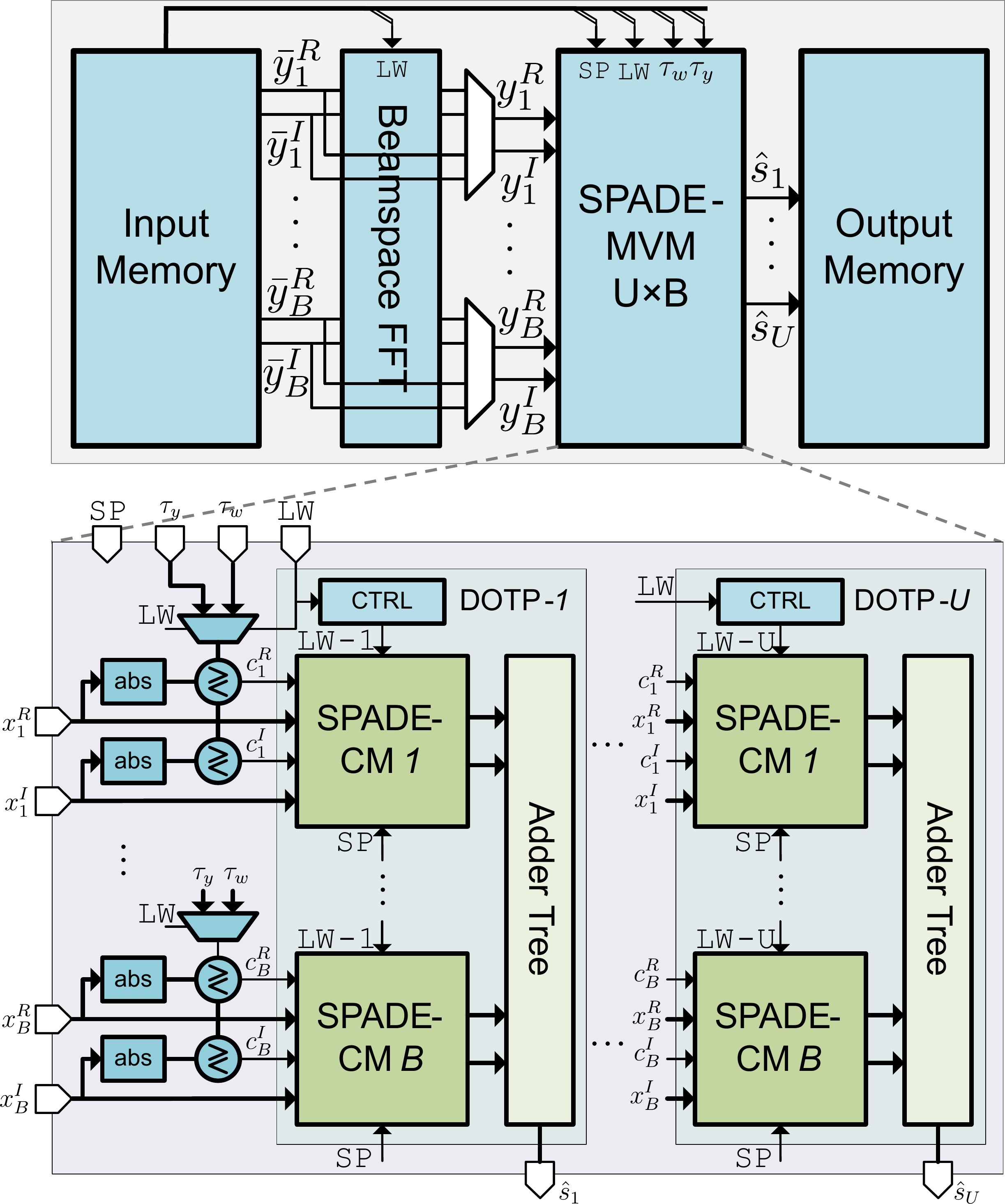}
\vspace{-0.15cm}
\caption{Top: high-level architecture; bottom: architecture of SPADE-MVM.}
\label{fig:top}
\end{figure}

The high-level architecture of our ASIC is illustrated at the top of \fref{fig:top} and consists of four main components: (i) an input SRAM, which stores  input test vectors, (ii) a fast Fourier transform (FFT) unit that transforms the antenna-domain received vectors into beamspace, (iii)  a SPADE-enabled matrix-vector multiplier (MVM), and (iv) an output SRAM that stores the results. 
The beamspace FFT is implemented using a fully-unrolled radix-4 architecture with low-resolution twiddle factors according to \cite{mirfarshbafan21}; this enables transforming an entire $B$-element antenna-domain vector into beamspace per clock cycle at a small area and power overhead. 
To compare the efficacy of beamspace vs.~antenna-domain processing, our architecture can be configured to operate either in antenna-domain mode, in which the beamspace FFT is turned off, or in beamspace mode, in which the beamspace FFT is active. 
In addition, when operating in beamspace mode, a save-power (SP) signal controls whether SPADE-based power saving is activated; this enables us to measure the impact of SPADE.

\begin{figure}[tp]
\centering
\includegraphics[width=0.9\columnwidth]{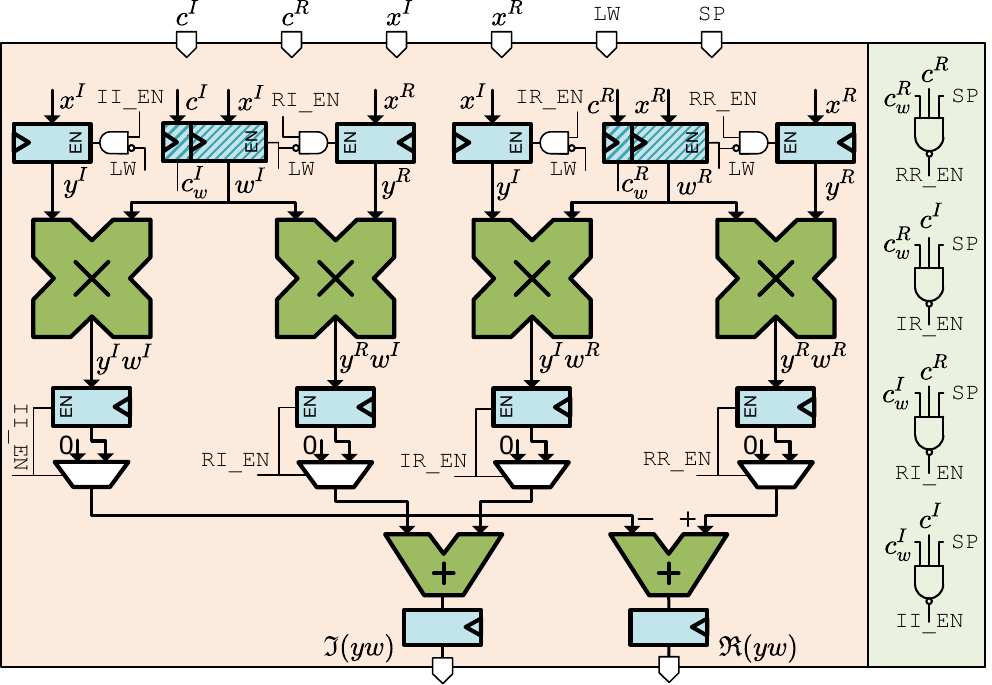}
\vspace{-0.2cm}
\caption{Detailed architecture of one SPADE complex-valued multiplier (CM).}\label{fig:spadecm}
\end{figure}


\begin{table*}[t]
	\centering
	\renewcommand{\arraystretch}{0.85}
	\begin{minipage}[c]{1\textwidth}
		\vspace{-0.1cm}
		\centering
		\caption{Comparison with state-of-the-art data detectors for massive MU-MIMO systems}
		\vspace{-0.1cm}
		\label{tbl:copmarison}
		\resizebox{0.7\textwidth}{!}{%
			\begin{tabular}{@{}lcc|ccccc@{}}
				\toprule
				~ & \multicolumn{2}{c|}{This work} &\cite{sirpac21} &  \cite{liu20} &  \cite{tang21}  &  \cite{prabhu17} &  \cite{Mahdavi20} \\
				\midrule
				Algorithm  & \multicolumn{2}{c|}{LMMSE-SPADE} & FLMMSE & RCG  & MPD & LMMSE & ZF-BS\\
				System dimension $B\times U$ & \multicolumn{2}{c|}{64 $\times$ 16} & 32 $\times$ 16 & 128 $\times$ 8 & 128 $\times$ 32  &  128 $\times$ 8 & 128 $\times$ 16\\
				Modulation [QAM] & \multicolumn{2}{c|}{16} & 16 & 64 & 256 & 256 & 16\\
				Channel scenario & LoS & NLoS & -- \footnotemark[1]  &  -- & --  & -- & -- \\
				Includes preprocessing & \multicolumn{2}{c|}{no} & no & no & no & yes & yes\\
				\midrule
				Technology [nm] &  \multicolumn{2}{c|}{22} & 65 & 65 & 40 & 28 & 28\\
				Core voltage [V] &  \multicolumn{2}{c|}{0.8} & 1.0 & 1.2 & 0.9 & 0.9 & -- \\
				Core area [$\text{mm}^2$] & \multicolumn{2}{c|}{2.3 \footnotemark[2]}   &  2.41 & 1.6 & 0.58 & 0.12  & --\\
				Max. clock frequency [MHz] & 720 & 600 & 312 & 500 & 425 & 300 & 560\\
				Max. throughput [Gbps] \footnotemark[3] & 46 & 39 & 9.98 & 1.0 & 1.38 & 0.15 & 2.24\\
				Max. clock frequency [MHz] (FBB) \footnotemark[4] & 920 & 890 & -- & -- & -- & 300 & -- \\
				Max. throughput [Gbps] \footnotemark[3] (FBB)  & 58.8 & 54 & -- & -- & -- & 0.15 & -- \\
				Power [mW]  & 544 \footnotemark[2]\footnotemark[5] & 570 \footnotemark[2]\footnotemark[5] & 290 & 120 & 220.6 & 18 & 251\\
				\midrule 
				Energy eff. [pJ/b] & 11.8 \footnotemark[5]  & 14.6 \footnotemark[5] & 29 & 120 & 160 & 120 & 112\\
				Area eff. [Gbps/$\text{mm}^2$] & 20 \footnotemark[5]  & 17 \footnotemark[5] & 4.15 & 0.63 & 2.38 & 1.25 & --\\
				\midrule
				Norm. energy eff. [pJ/b] \footnotemark[6]\footnotemark[7] & 11.8  & 14.6 & 12.6 & 36.1 &  34.6 & 148 & 69.5\\
				Norm. area eff. [Gbps/$\text{mm}^2$] \footnotemark[6]\footnotemark[7]  & 20  & 17 &  53 & 7.8 & 28.6 & 1.28 & --\\
				
				\bottomrule
			\end{tabular}
			
		}\vspace{-0.1cm}
			\newcommand{\footmark}[1]{$^{\footnotesize \textnormal{#1}}$}
			\footnotetext{
			\footmark{a}not reported information indicated by -- ; 
			\footmark{b}excluding test memories; 
			\footmark{c}computed with the max. reported clock frequency for $16$-QAM in all designs to isolate the effect of modulation order;
			\footmark{d}measured with forward body biasing;
			\footmark{e}measured without body biasing;
			\footmark{f}technology normalized to 22\,nm at nominal core supply where the throughput is scaled by $s$, the area by $1/s^2$, and the power by $1/(V^2)$, where $s$ is the ratio of technology nodes and $V$ is the ratio of core voltages \cite{Rabaey_book};
			\footmark{g}area and power scaled by $(16/U)^2$ according to \cite{prabhu17, Mahdavi20} and throughput scaled by $16/U$, except for \cite{Mahdavi20} whose area and power scales with $BU$.}
		
	\end{minipage}
\vspace{-0.4cm}	
\end{table*}

\subsection{SPADE-MVM Architecture}
The top-level architecture of the SPADE-MVM is depicted at the bottom of \fref{fig:top} and consists of $U=16$ dot-product (DOTP) modules, each consisting of $B=64$ SPADE complex-valued multipliers (CMs), whose internal architecture is shown in \fref{fig:spadecm}, and a $B$-input adder tree consisting of $\text{log}_2(B)$ adder layers, with pipeline registers after every two layers. 
Each of the dot product modules computes one entry of $\bW\bmy$. 
SPADE-MVM receives a $B$-element complex-valued vector per clock cycle through its input ports. During a $U$-cycle weight-loading phase, indicated by raising a load-weight (LW) signal, a new equalization matrix $\bW$ is loaded into the registers of the SPADE-CMs.  When LW is low, the incoming vector $\bmy$ is multiplied to the rows of the stored $\bW$ simultaneously, performing one equalization operation $\bW\bmy$ per clock cycle.
Furthermore, if LW is asserted, the absolute value of the real and imaginary parts of the input signals are compared to the w-threshold $\tau_w$ and otherwise, they are compared to the y-threshold $\tau_y$. The resulting comparison bits are broadcast to the SPADE-CMs, where they are used for adaptive power~saving.

\subsection{SPADE-CM Architecture}

\fref{fig:spadecm} details the SPADE-CM architecture which contains four real-valued multipliers and two adders. The pipelining registers not only shorten the critical path and reduce glitching activity, but also provide a mechanism to conditionally mute each of the four real-valued multipliers by freezing their inputs. 
Two of the registers (cf.~the hatched pattern in \fref{fig:spadecm}) store the entries of the equalization matrix along with the w-threshold comparison bits ($c_w^R$ and $c_w^I$).
The other four input registers hold the real/imaginary parts of the input signal. If the save-power (SP) signal is asserted, and both the w- and y-threshold comparison bits corresponding to a particular register are set, then that register is disabled to mute switching activity, which reduces dynamic power consumption. At the input of the adders, the multiplexer selects zero if the preceding multiplier was muted. 
Depending on the channel conditions (e.g., depending on how sparse~$\bH$ is) and the instantaneous receive vector, different subsets of multipliers within SPADE-CMs are muted; this results in \emph{adaptive} savings in dynamic power.


\section{ASIC Implementation Results} 
\label{sec:impl}

\begin{figure}[t]
	\centering
	\includegraphics[width=0.42\columnwidth]{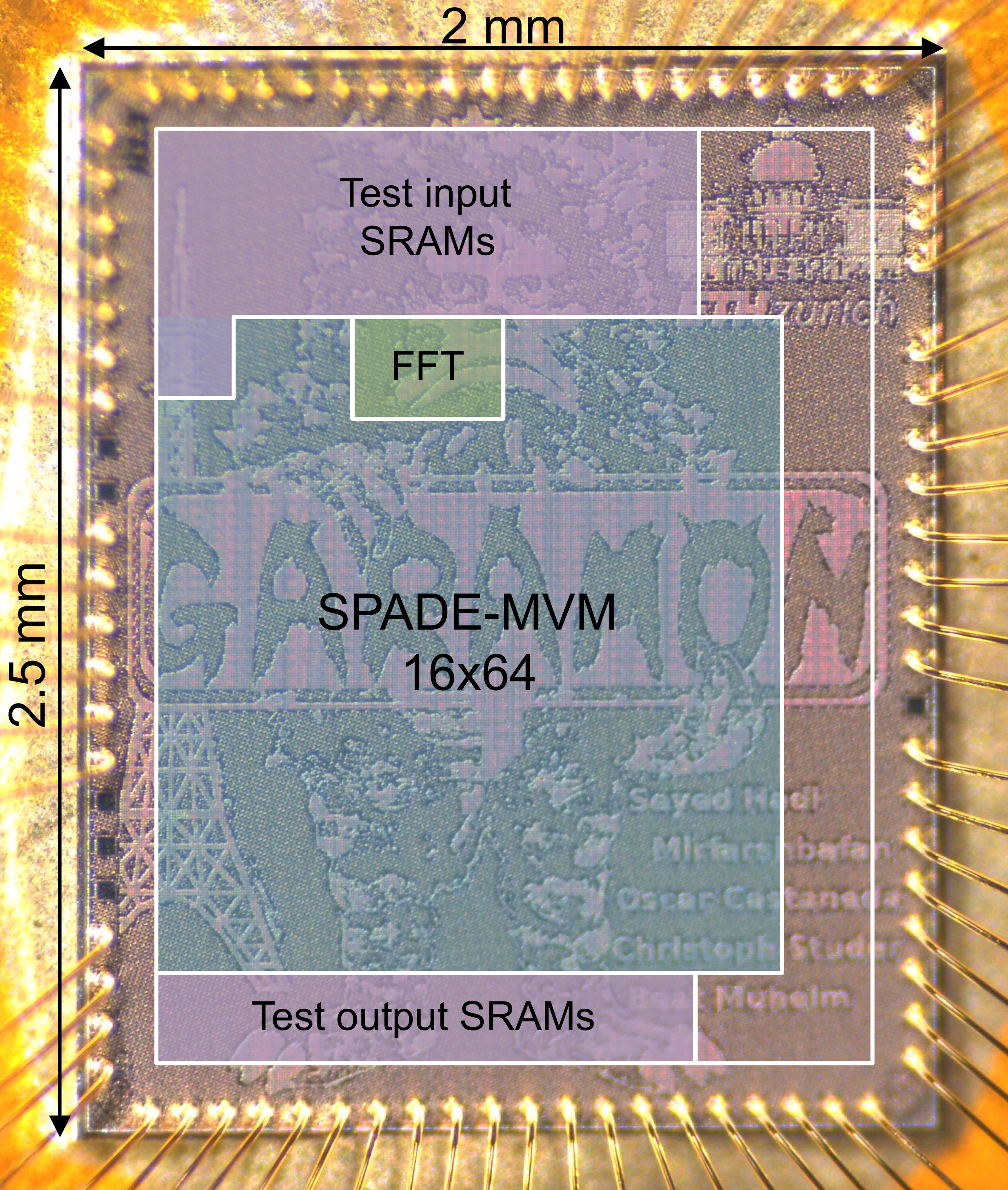}
	\hfill
	\includegraphics[width=0.43\columnwidth]{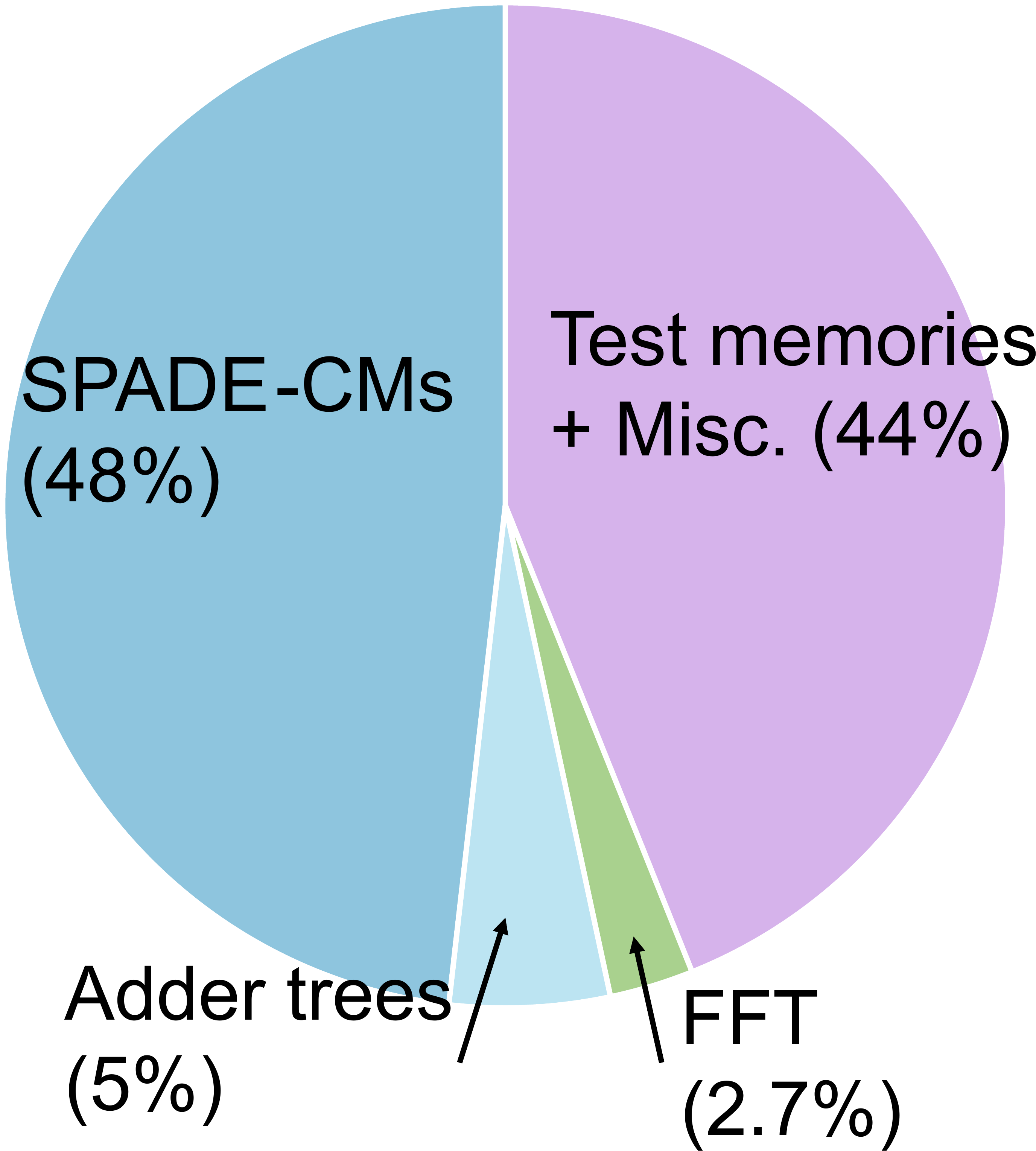}
	\vspace{-0.15cm}
	\caption{(left) Micrograph of the fabricated ASIC with the key blocks highlighted; (right) area breakdown of all active circuitry.}
	\label{fig:chip_micrograph}
\end{figure}

\begin{figure}[t]
	\centering
	\vspace{-0.35cm}
	\subfigure[]
	{\includegraphics[width=0.455\columnwidth]{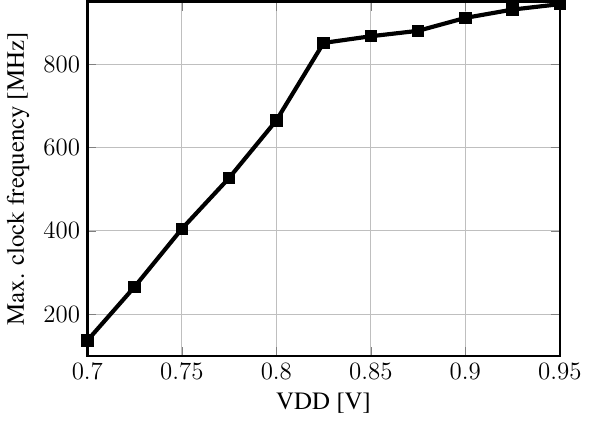}\label{fig:vfs}}
	\hfill
	\subfigure[]
	{\includegraphics[width=0.5\columnwidth]{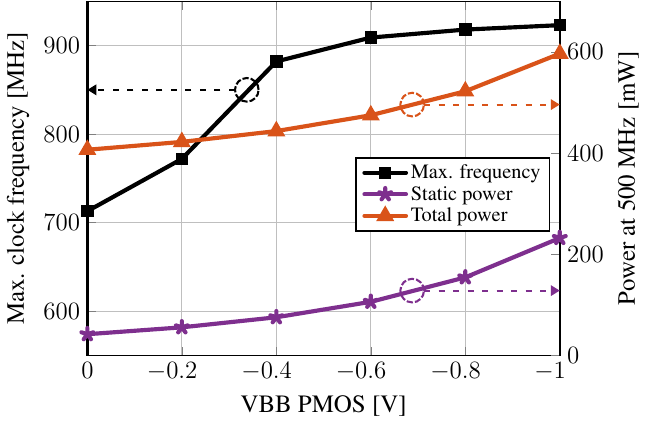}\label{fig:bb}}
	\vspace{-0.15cm}
	\caption{ASIC measurement results:  (a) voltage-frequency scaling (VFS); (b) clock frequency and equalizer power vs. body biasing voltage; power is measured in LMMSE-SPADE mode with LoS channels for $U=16$.}
	\vspace{-0.0cm}
	\label{fig:measurements}
\end{figure}

\begin{figure}[t]
\centering
\includegraphics[width=0.98\columnwidth]{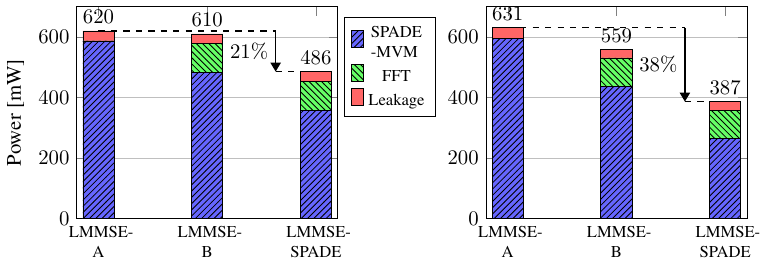}
\vspace{-0.25cm}
\caption{Power breakdown at $500$\,MHz for the three operation modes of our ASIC with inputs from non-LoS (left) and LoS channels (right) for $U=16$.}
\label{fig:power_brk}
\end{figure}

A micrograph of the fabricated ASIC along with an area breakdown is shown in \fref{fig:chip_micrograph}.
\fref{fig:vfs} shows the impact of the core voltage (VDD) on the maximum clock frequency and \fref{fig:bb} shows the effect of forward body biasing (FBB) on the maximum clock frequency and the power consumption at a clock frequency of $500$\,MHz. For simplicity, in \fref{fig:bb}, only the PMOS body biasing voltage is varied and the NMOS body bias is fixed to $0.4$\,V. We observe that $1.0$\,V of PMOS forward body biasing, results in a $29$\% increase in the maximum clock frequency and a $46$\% increase in the total power consumption, which is due to the increase in leakage power.
\fref{fig:power_brk} shows the power breakdown of our ASIC's main modules in three modes: (i) LMMSE-A, in which the beamspace FFT is turned off and antenna-domain equalization is performed, (ii) LMMSE-B, in which beamspace equalization is performed but without SPADE-based power saving (no multiplications are muted), and (iii) LMMSE-SPADE, which performs beamspace equalization with SPADE-based power saving. LMMSE-SPADE provides $21$\% and $38$\% power savings with respect to LMMSE-A under non-LoS and LoS conditions, respectively.

\fref{tbl:copmarison} compares the key characteristics of our fabricated ASIC with that of state-of-the-art massive MU-MIMO data detectors. We scale the reported metrics as detailed in footnotes $^{\footnotesize \textnormal{f}}$ and $^{\footnotesize \textnormal{g}}$. Our ASIC achieves similar or superior normalized energy- and area-efficiency compared to \revision{state-of-the-art hardware implementations.} Even though the designs in~\cite{liu20, tang21} achieve competitive energy and area efficiency, their BER is inferior in mmWave channels (cf. \fref{fig:BER} and \cite{tang21}). The design from~\cite{sirpac21} achieves better area-efficiency by utilizing extremely low-resolution equalization weights and input vectors. This design, however, would require \revision{a more complicated preprocessing engine} than our design. Finally, we emphasize that our ASIC achieves up to $58.8$\,Gbps at only $833$\,mW with body biasing, which is, to our knowledge, the highest 16-QAM throughput reported in the literature.


\section{Conclusions}
\label{sec:conclusion}

We have presented the first architecture and ASIC of a mmWave massive MU-MIMO equalizer capable of both antenna-domain and beamspace equalization with adaptive power-saving capabilities. In beamspace equalization mode, our ASIC \revision{is able to use} SPADE~\cite{mirfarshbafan21b} to adaptively reduce the dynamic power consumption. 
Measurement results have revealed that, despite the overhead of the \revision{necessary} beamspace FFT, beamspace equalization with SPADE enables up to $38$\% power savings compared to antenna-domain equalization. 
Furthermore, our ASIC achieves a record throughput of $58.8$\,Gb/s with body biasing at \revision{a} better or similar normalized energy- and area-efficiency compared to state-of-the-art designs. 

\newpage


\end{document}